\documentclass[twocolumn,prb,showpacs,floatfix]{revtex4}
\usepackage{graphicx}
\usepackage{amsfonts}
\usepackage{amssymb}
\usepackage{bm}

\newcommand{\be}{\begin{equation}}
\newcommand{\ee}{\end{equation}}
\newcommand{\bea}{\begin{eqnarray}}
\newcommand{\eea}{\end{eqnarray}}
\newcommand{\zncu}{ZnCu$_3$(OH)$_6$Cl$_2$}

\begin{document}

\title{Kagome lattice antiferromagnets and Dzyaloshinsky-Moriya interactions}

\author{Marcos Rigol}
\affiliation{Department of Physics and Astronomy,
University of Southern California, Los Angeles, California 90089, USA}
\author{Rajiv R.~P.~Singh}
\affiliation{Department of Physics, University of California, Davis,
California 95616, USA}

\date{\today}

\pacs{75.10.Jm,05.50.+q,05.70.-a}
% 75.10.Jm Quantized spin models

\begin{abstract}
We study the consequences of in-plane ($D_p$) and out-of-plane ($D_z$) 
Dzyaloshinsky-Moriya (DM) interactions on the thermodynamic properties 
of spin-$\frac{1}{2}$ Heisenberg model on the kagome lattice using 
numerical linked cluster expansions and exact diagonalization, 
and contrast them with those of other perturbations such as exchange 
anisotropy and dilution. We find that different combinations of the DM 
anisotropies lead to a wide variety of thermodynamic behavior, which
are quite distinct from those of most other perturbations. We argue 
that the sudden upturn seen experimentally in the susceptibility of 
the material \zncu\ can be understood in terms of Dzyaloshinsky-Moriya 
anisotropies with $D_p$$>$$|D_z|$. We also show that the measured specific 
heat of the material puts further constraints on the allowed DM parameters.
\end{abstract}

\maketitle

\section{Introduction \label{introduction}}

The possible realization of exotic states of matter has always been
at the forefront of research interest in condensed matter physics. 
One class of states, which has received considerable interest over 
the last few decades, consists of quantum spin liquids.\cite{anderson73} In 
these states, no magnetic order or other symmetry breaking occurs as 
the temperature is lowered, while the system may or may not exhibit 
a gap in its excitation spectra. Two variants of the quantum 
spin liquid have received particular attention recently: (i) a 
topological spin liquid, which has a spin gap and a topological 
order,\cite{kivelson87,moessner01} and (ii) an algebraic spin liquid, 
where there is no spin gap and spin-spin correlations decay as a 
power law.\cite{hastings01,hermele05} Kagome lattice antiferromagnets 
are potential candidates for both these types of spin liquids.

\begin{figure}[!htb]
\begin{center}
  \includegraphics[scale=.3,angle=0]{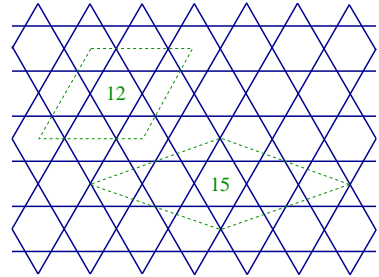}
\end{center}
\vspace{-0.5cm}
\caption{\label{KagomeLattice}
(Color online) The kagome lattice. The embedded finite clusters
are used (with periodic boundary conditions) in the ED study of DM 
interactions.}
\end{figure}

Recently, the newly synthesized herbertsmithite\cite{shores05} \zncu\ 
has brought tremendous excitement to the field. For this rare mineral, 
in which the spin-$\frac{1}{2}$ copper atoms form a kagome lattice, 
like the one depicted in Fig.\ \ref{KagomeLattice}, no magnetic order 
is observed down to 50 mK (Refs.\ \onlinecite{helton07,ofer07,mendels07,imai07}) 
even though the exchange constant is approximately 
170 K.\cite{rigol07}

The spin-$\frac{1}{2}$ kagome lattice Heisenberg model (KLHM) has been
extensively studied using series expansions and exact diagonalization 
of finite-size periodic clusters.\cite{zeng90,singh92,leung93,elstner94,lecheminant97,waldtmann98,sindzingre00,misguich05}
Exact diagonalization (ED) studies suggest that this model does not exhibit any magnetic order but 
possibly has a small spin gap $\sim$$J/20$. In addition, in finite systems, 
it has been found that there are a large number of singlet states below the 
spin gap and that their number grows with the system size,\cite{waldtmann98,mila98} 
which indicates that in the thermodynamic limit, nonmagnetic excitations may 
develop a continuum beside the ground state.

Considering the above theoretical results, the experimental behavior of the 
recently synthesized kagome systems, ZnCu$_3$(OH)$_6$Cl$_2$, is highly
unexpected.\cite{helton07,ofer07} At high temperatures, the inverse 
susceptibility data was found to obey a Curie-Weiss law, with an effective 
Curie-Weiss constant of about 300 K. However, no spin gap was seen either 
in the susceptibility, the specific heat, the neutron spectra, or in the 
nuclear spin-lattice relaxation $T_1$, down to temperatures below 100 mK. 
In addition, at the lowest temperatures, the susceptibility saturates 
to very high values and the specific heat shows power-law behavior in 
temperature. The latter is suppressed by magnetic fields, showing it to be 
magnetic in origin. 

%%%%%%%%%%%%%%  FIGURE  %%%%%%%%%%%%%%%%%%%%%%%%%%%%%%%%%%%%%%%%%%%%%%
%%%%%%%%%%%%%%%%%%%%%%%%%%%%%%%%%%%%%%%%%%%%%%%%%%%%%%%%%%%%%%%%%%%%%%%%

A possible interpretation of the substantial rise in the susceptibility
seen experimentally at low temperatures is that it is due to impurity 
spins outside the kagome planes, possibly caused by substitutions of 
non magnetic Zn sites with Cu.\cite{ran06} This idea has been reinforced 
in recent numerical analysis of Misguich and Sindzingre.\cite{misguich07} 
They find that the deviation of the experimental data from the kagome 
lattice Heisenberg model at least down to $T$$\approx$40 K can be understood 
in terms a small concentration of impurities provided the impurity 
contribution also has a weak ferromagnetic Curie-Weiss constant. We, on 
the other hand, have argued that the sharp rise in the susceptibility seen 
experimentally is intrinsic and it is related to the presence of Dzyaloshinsky-Moriya 
(DM) interactions.\cite{rigol07} The latter is consistent with the experimental 
observation that the muon shift $K$ tracks the bulk susceptibility $\chi$.

The two different explanations can be distinguished by studying 
single crystals, whereby the susceptibilities along different 
crystallographic axes can be investigated. The explanations based on DM 
interactions lead to substantial temperature dependent anisotropy in the 
susceptibility, whereas the impurity contributions should be isotropic. 
In the absence of single crystals, recent NMR work of Imai 
{\it et al.}\cite{imai07} may already provide some resolution of the 
issue. Imai {\it et al.} find that the NMR spectra progressively broaden 
as one goes to lower temperatures. While the temperature dependence 
of the median of the broadened spectra resembles the sharp upturn seen 
in bulk susceptibility measurements, the spectra at the edges do not show 
this upturn. Rather, they show a saturation and eventually a downturn with 
lowering of temperature as expected in antiferromagnets, when short range 
order sets in. Indeed, we will see here that for certain choices of the DM 
parameters, susceptibilities along different crystallographic axes show the 
two different behaviors observed by Imai {\it et al}. Thus, one interpretation 
of the NMR experiments is that they are observing the susceptibility along 
different axes, due to the different alignment of the powders with respect 
to the applied field.

In this work, we further develop the DM theory by
presenting a detailed study of the effects of in-plane 
($D_p$) and out-of-plane ($D_z$) Dzyaloshinsky-Moriya interactions, 
as well as other perturbations such as easy-plane and easy-axis exchange 
anisotropies and quenched dilution, on the thermodynamic properties 
of the KLHM. We use ED and the triangle-based numerical linked cluster 
method (NLC)\cite{nlc} to compute the uniform susceptibility, entropy, 
and specific heat. We find that unlike exchange anisotropy or dilution, 
the effects of DM interactions on the susceptibility can set in quite 
abruptly as a function of temperature. Furthermore, different choices of the
DM parameters can lead to a wide range of behavior for the susceptibility.
In particular, the abrupt upturn in the susceptibility seen experimentally 
in \zncu\  around 75 K can be understood in terms of Dzyaloshinsky-Moriya 
anisotropies when $D_p$$>$$|D_z|$.

Also comparing the experimental specific heat data with their 
theoretical results, Misguich and Sindzingre\cite{misguich07}
concluded that the measured entropy in the materials \zncu\ at $T/J=0.06$
is much lower than for the pure KLHM, and that the very low temperature
specific heat may be dominated by impurities. We discuss the role of DM
interactions in the entropy and specific heat of the material. In
general, DM interactions should lower the entropy as they reduce the
manifold of classical ground states. We find that the entropy is reduced
primarily due to $D_z$, whereas $D_p$ has a very small effect on it.
Thus, combining the experimental results on the susceptibility and
specific heat, we believe that the most likely DM parameters for the material 
are in the range, $D_p/J\approx 0.2-0.3$, $|D_z|/J\approx 0.1$. 
However, these results may change a little if impurity effects are 
included in the analysis.

The results of Misguich and Sindzingre\cite{misguich07} further 
substantiate our earlier assertion that the pure KLHM has an extended 
crossover regime,\cite{rigol07} where the susceptibility grows as 
a power law in inverse temperature and the specific heat or entropy 
is sublinear in temperature. From their numerical results, the 
power law in the susceptibility may extend over a full decade in 
temperature, $0.1<T/J<1$. We would like to reiterate that this 
crossover regime is important to address theoretically and may well 
be relevant to the properties of the real material. 

This exposition is organized as follows. In Sec.\ \ref{susceptibility}, 
we discuss the effects of Dzyaloshinsky-Moriya and exchange
anisotropy on the susceptibility and compare them with experiments on
\zncu. In Sec.\ \ref{entropycv}, we study the consequences of the 
above mentioned perturbations and of quenched dilution on the entropy 
and specific heat and also discuss the constraints on the DM parameters 
for \zncu\ that the specific heat (entropy) measurements introduce.
Finally, the conclusions are presented in Sec.\ \ref{conclusions}.

\section{Uniform susceptibility \label{susceptibility}}

In this section, we study the effects of different perturbations on the 
uniform susceptibility of the spin-$\frac{1}{2}$ Heisenberg model. 
In a magnetic field ${\bf h}$, the Hamiltonian of this model can be written as
\begin{equation}
{\cal H}=J\sum_{\langle i,j\rangle} {\bf S}_i {\bf S}_j 
+{\cal H}_{pert}
- g\mu_B \sum_i {\bf h\ S}_i.
\end{equation}
where $J$ is the exchange coupling, $g$ is the $g$ factor (assumed to be
isotropic), $\mu_B$ is the Bohr magneton, and 
${\cal H}_{pert}$ represents the various perturbations. In what follows, 
we set $J=1$ and $g\mu_B=1$. In the sum, $\langle i,j\rangle$ means that 
only nearest neighbor interactions are considered.

The uniform susceptibility per spin is then given by 
\begin{equation}
\chi_\alpha=\frac{T}{N}\left.\frac{\partial^2 \ln{Z}}
{\partial h_\alpha^2}\right\vert_{{\bf h}=0},
\label{suscept}
\end{equation}
where $Z$ is the partition function, $T$ the temperature, 
$N$ the number of lattice sites, and $\alpha=x,y,z$. The molar 
susceptibility $\chi_{molar}$, measured experimentally, is related 
to our susceptibility per spin by the relation $\chi_{molar}= C \chi$, 
where the constant $C=N_A g^2\mu_B^2/kJ=0.3752\ g^2/J$ in cgs units.

%%%%%%%%%%%%%%  FIGURE  %%%%%%%%%%%%%%%%%%%%%%%%%%%%%%%%%%%%%%%%%%%%%%
\begin{figure}[!htb]
\begin{center}
  \includegraphics[scale=.4,angle=0]{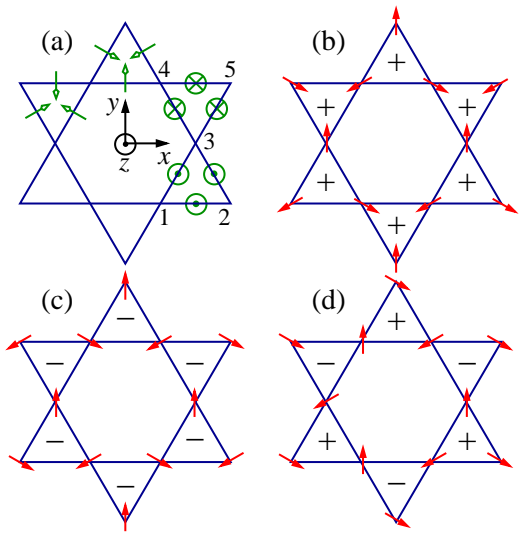}
\end{center}
\vspace{-0.5cm}
\caption{\label{ClassicalKagome}
(Color online) (a) Orientation of the in-plane and out-of-plane DM
interactions. [(b)-(d)] Three possible classical ground states of the 
kagome lattice. In each case, the system has (b) positive chirality, 
(c) negative chirality, and (d) both positive and negative chiralities, 
as indicated by the signs inside the triangles.}
\end{figure}
%%%%%%%%%%%%%%%%%%%%%%%%%%%%%%%%%%%%%%%%%%%%%%%%%%%%%%%%%%%%%%%%%%%%%%%%

In a kagome lattice (which we assume lies 
in the $x$-$y$ plane), both out-of-plane ($D_z$) and in-plane ($D_p$) 
DM terms are allowed,\cite{dzyaloshinsky58,moriya60,elhajal02,harris06,yamabe06}
\begin{equation}
{\cal H}_{DM}=\sum_{\langle i,j\rangle}  D_z ({\bf S_i}\times{\bf S_j})_z
+{\bf D_p}\cdot ({\bf S_i}\times{\bf S_j}),
\label{DMH}
\end{equation}

Since the DM terms break spin rotational symmetry, we need to calculate 
separately the susceptibility with field along $z$ ($\chi_z$) and in the 
$x$-$y$ plane ($\chi_p$). The powder susceptibility $\chi_a$ is given by 
$\chi_a={1\over 3}(2\chi_p+\chi_z)$. 

With DM interactions, the different $S^z$ sectors become coupled so that we 
are able to do NLC calculations only up to six triangles.\cite{nlc} Unfortunately, 
the convergence of the KLHM with additional DM anisotropy is poor, and 
having only very few terms for the NLC expansion does not allow us to perform 
extrapolations. Hence, we turn to ED of clusters with 12 and 15 sites 
(Fig.\ \ref{KagomeLattice}), and periodic boundary conditions, to study the 
effects of DM interactions. In all our plots, we include both the results 
for 12 and 15 sites. The region where they agree gives an idea of the 
temperature range where finite-size effects are small and one can be confident 
of the ED results. 

\subsection{$D_z\neq0,\ D_p=0$}

We first consider the case of a pure out-of-plane DM interaction 
($D_z\neq0,\ D_p=0$). The sign of the $D_z$ term alternates between 
the up- and down-pointing triangles of the kagome lattice. It can be 
set by demanding that for the up pointing triangle shown in 
Fig.\ \ref{ClassicalKagome} with corners 1-2-3, a positive 
$D_z$ multiplies $({\bf S_1}\times{\bf S_2})_z$, and for the 
down-pointing triangle 3-4-5, a negative $D_z$ multiplies 
$({\bf S_4}\times{\bf S_5})_z$.

The effect of a pure out-of-plane DM interaction is to favor the spins
to lie in the $x$-$y$ plane. In that sense, it acts like an {\it easy-plane
exchange anisotropy}. In addition, the sign of $D_z$ breaks the chiral 
symmetry of the KLHM. In the classical limit, the ground state of the
Heisenberg antiferromagnet is highly degenerate. All states satisfying 
the ``120$^\mathrm{o}$'' rule in each triangle minimize the 
energy. In Figs.\ \ref{ClassicalKagome}(b)-\ref{ClassicalKagome}(d), 
we show three possible N\'eel states of the KLHM, where spins lie in 
the plane and which have positive, negative, and mixed chiralities, 
respectively. Once the $z$ component of the DM interaction is introduced, 
the degeneracy between the states in Fig.\ \ref{ClassicalKagome} is lifted. 
For $D_z>0$, the state with negative chirality [Fig.\ \ref{ClassicalKagome}(c)] 
is favored, while for $D_z<0$, the state with positive chirality 
[Fig.\ \ref{ClassicalKagome}(b)] is the one that minimizes the energy.

%%%%%%%%%%%%%%  FIGURE  %%%%%%%%%%%%%%%%%%%%%%%%%%%%%%%%%%%%%%%%%%%%%%
\begin{figure}[!t]
\begin{center}
  \includegraphics[scale=.6,angle=0]{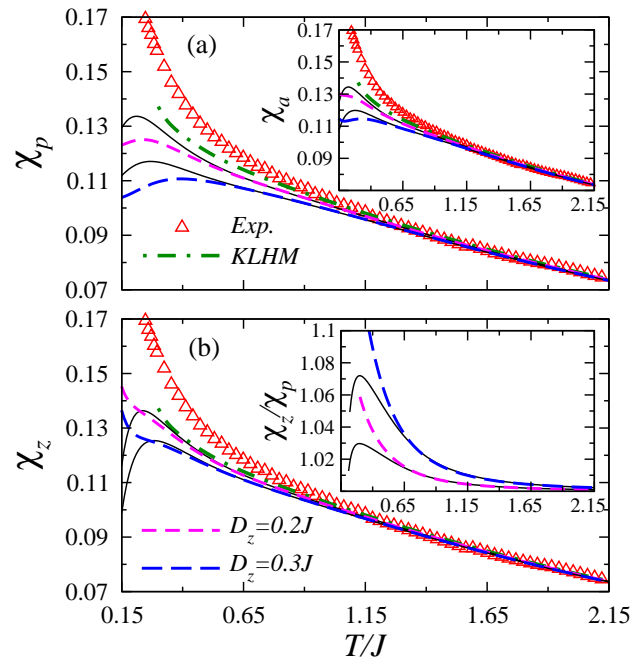}
\end{center}
\vspace{-0.5cm}
\caption{\label{DM_z_Susceptibility}
(Color online) Temperature dependence of the (a) in-plane and (b) out-of-plane 
susceptibilities in the presence of a pure $D_z$ anisotropy. They are compared 
with $\chi$ for the pure KLHM [computed using the triangle based NLC
expansion considering up to eight triangles (Ref.\ \onlinecite{nlc})] and with 
the experimental results (Refs.\ \onlinecite{helton07} and \onlinecite{ofer07}) 
translated to our notation (Ref.\ \onlinecite{experiments}). 
In the inset in (a), we show the powder susceptibility and in the inset in (b) 
the anisotropy. In all plots of DM calculations, thick (thin) lines show the 
ED results of the 15 (12) site cluster.}
\end{figure}
%%%%%%%%%%%%%%%%%%%%%%%%%%%%%%%%%%%%%%%%%%%%%%%%%%%%%%%%%%%%%%%%%%%%%%%%

Returning to our spin-half quantum model, in Fig.\ \ref{DM_z_Susceptibility} we
show the in-plane and out-of-plane susceptibilities as a function of 
temperature for two different strengths of $D_z$. (If only $D_z$ 
is present, its sign is irrelevant for the thermodynamic quantities 
we study in this paper.) We compare these results with the ones 
obtained for the pure KLHM\cite{nlc} and the ones measured 
experimentally for \zncu.\cite{helton07,ofer07}\  Two features are apparent 
in these plots, which can be related to the increase of in-plane 
correlations between spins, as expected from classical considerations. 
(i) The $D_z$ term suppresses both $\chi_p$ and $\chi_z$ 
with respect to the KLHM result and (ii) $\chi_z$ becomes larger than 
$\chi_p$. Since the experimental result for the powder susceptibility
is larger than the one for the KLHM, it is evident
[inset in Fig.\ \ref{DM_z_Susceptibility}(a)] that a $D_z$ 
term alone cannot explain the experiments. The anisotropy 
produced by $D_z$ in the susceptibilities is shown in the inset in 
Fig.\ \ref{DM_z_Susceptibility}(b).

There is another feature in the plots in Fig.\ \ref{DM_z_Susceptibility}
that is noticeable: finite-size effects set in at higher temperatures 
for $\chi_p$ than for $\chi_z$. This confirms that the planar ($XY$) 
correlations are longer ranged than the $ZZ$ correlations.

\subsection{$D_p\neq0,\ D_z=0$}

We now analyze an in-plane DM interaction $D_p$. It is perpendicular to 
the bonds and points inward toward the center of the 
triangles.\cite{harris06,yamabe06} This is shown by the arrows in 
Fig.\ \ref{ClassicalKagome}(a). (A different scenario for $D_p$ is discussed
in the Appendix.) $D_p$ breaks the rotational
symmetry around the $c$ axis. It also favors classical spin configurations with 
a finite $z$ component, producing weak ferromagnetism.\cite{elhajal02,harris06} 
In that sense, $D_p$ can act like an {\it easy-axis exchange anisotropy}.

%%%%%%%%%%%%%%  FIGURE  %%%%%%%%%%%%%%%%%%%%%%%%%%%%%%%%%%%%%%%%%%%%%%
\begin{figure}[!hb]
\begin{center}
  \includegraphics[scale=.6,angle=0]{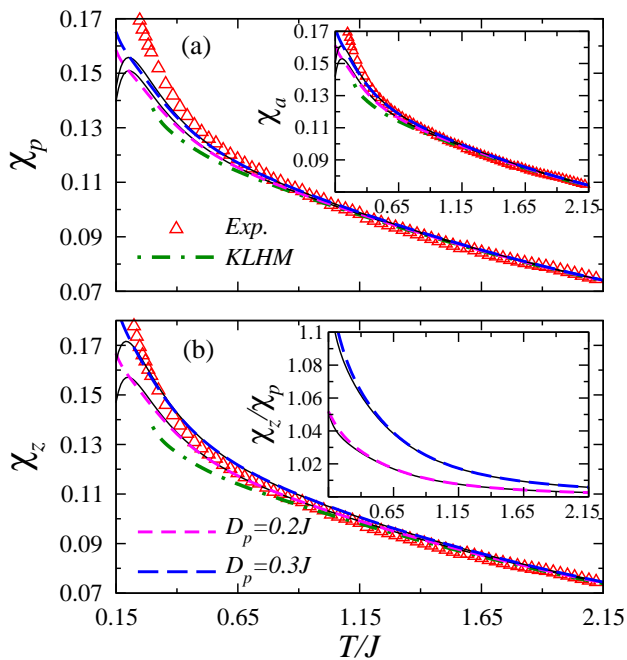}
\end{center}
\vspace{-0.5cm}
\caption{\label{DM_p_Susceptibility}
(Color online) Temperature dependence of (a) the in-plane and (b) out-of-plane 
susceptibilities in the presence of a pure $D_p$ anisotropy. They are compared 
with $\chi$ for the pure KLHM (Ref.\ \onlinecite{nlc}) and with the experimental 
results (Refs.\ \onlinecite{helton07} and \onlinecite{ofer07}). In the inset in 
(a), we show the powder susceptibility, and in the inset in (b) the anisotropy. 
In all plots of DM calculations, thick (thin) lines show the ED results of 
the 15 (12) site cluster.}
\end{figure}
%%%%%%%%%%%%%%%%%%%%%%%%%%%%%%%%%%%%%%%%%%%%%%%%%%%%%%%%%%%%%%%%%%%%%%%%

In Fig.\ \ref{DM_p_Susceptibility}, we depict the $x$-$y$ and $z$ 
susceptibilities as a function of the temperature for two different 
strengths of $D_p$.\cite{sign} As for $D_z$, we compare our results 
with the ones obtained for the pure KLHM\cite{nlc} and the ones 
measured experimentally.\cite{helton07,ofer07} We find that a $D_p$ 
term enhances both $\chi_p$ and $\chi_z$ with respect to the KLHM result, 
and that $\chi_z$ becomes larger than $\chi_p$. This can be understood
considering that $D_p$, when trying to produce canting, competes with 
the Heisenberg terms and reduces the in-plane spin-spin correlations.

The enhancement of $\chi_p$ and $\chi_z$ seen in 
Fig.\ \ref{DM_p_Susceptibility} shows that a pure $D_p$ term 
can explain the upturn seen experimentally for the susceptibility. 
This can be seen in the plots of the powder susceptibilities 
presented in the inset in Fig.\ \ref{DM_p_Susceptibility}(a). 
Interestingly, the anisotropies produced by $D_z$ and $D_p$ 
are very similar, as can be concluded by comparing the inset in 
Fig.\ \ref{DM_p_Susceptibility}(b) with the one in 
Fig.\ \ref{DM_z_Susceptibility}(b). They both generate 
$\chi_z>\chi_p$, and $\chi_z/\chi_p$ are of the same order 
(at least for the intermediate temperatures studied here) 
when $D_z$ and $D_p$ are of the same order. 

A remarkable difference between the effect of a $D_z$ term 
and the effect of a $D_p$ term in our finite cluster calculation 
of the susceptibilities is that in the latter, finite-size effects 
are very small as compared to the former one. This implies that all 
correlations are much weaker in the presence of a $D_p$ term than 
in the presence of a $D_z$ term.

\subsection{$D_p\neq0,\ D_z\neq0$, and the experiments}

\begin{figure*}[!h]
\begin{center}
  \includegraphics[scale=.55,angle=0]{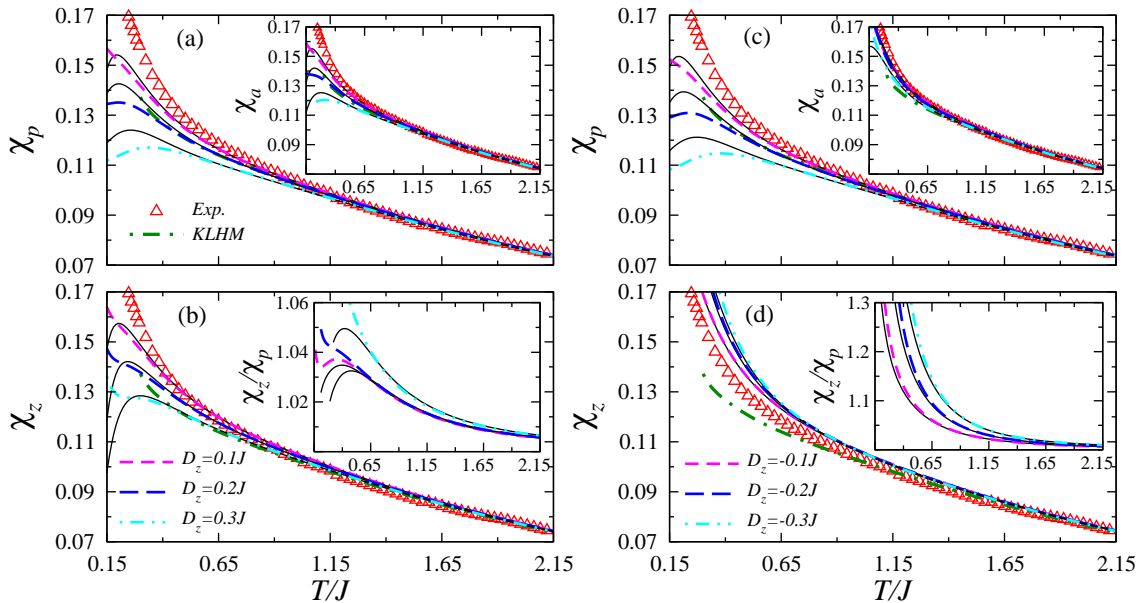}
\end{center}
\vspace{-0.5cm}
\caption{\label{DM_pgtz_Susceptibility}
(Color online) Temperature dependence of the [(a) and (c)] in-plane and 
[(b) and (d)] out-of-plane susceptibilities in the presence of an in-plane 
$D_p=0.3J$ anisotropy and different values and signs of the out-of-plane 
anisotropy. These susceptibilities are compared with $\chi$ for the pure 
KLHM (Ref.\ \onlinecite{nlc}) and with the experimental 
results (Refs.\ \onlinecite{helton07} and \onlinecite{ofer07}). 
In the inset in (a) and (c), we show powder susceptibilities, and in the 
inset in (b) and (d) the anisotropies. In all plots of DM calculations, 
thick (thin) lines show the ED results of the 15 (12) site cluster.}
\end{figure*}

Once $D_p\neq 0$, in general, thermodynamic quantities for $D_z>0$ start to 
differ from the ones for $D_z<0$. In Fig.\ \ref{DM_pgtz_Susceptibility}, we 
have plotted the in-plane and out-of-plane susceptibilities for a fixed value 
of $D_p$ while increasing $|D_z|$, with $D_z>0$ [(a) and (b)] and $D_z<0$ 
[(c) and (d)]. Comparing Fig.\ \ref{DM_pgtz_Susceptibility}(a) with 
Fig.\ \ref{DM_pgtz_Susceptibility}(c) one can see that the increase of $|D_z|$
produces the same effect in $\chi_p$ no matter the sign of $D_z$. For both
$D_z>0$ and $D_z<0$, the in-plane susceptibility is suppressed with respect to
its value for $D_p\neq 0, D_z=0$. On the other hand, the effect produced by 
the increase of $|D_z|$ on $\chi_z$ is very different depending on the sign 
of $D_z$. Figure \ref{DM_pgtz_Susceptibility}(b) shows that for $D_z>0$, the 
increase of $D_z$ suppresses $\chi_z$ with respect to its value for 
$D_p\neq 0,\ D_z=0$. As seen in Fig.\ \ref{DM_pgtz_Susceptibility}(d), 
the opposite occurs if $D_z<0$. The increase of $|D_z|$ enhances 
$\chi_z$ with respect to its value for $D_p\neq 0,\ D_z=0$.

Given the above results for the in-plane and out-of-plane susceptibilities
in the presence of $D_p$ and $D_z$, one can then understand the behavior 
of the powder averages, which is presented in the insets of
Figs.\ \ref{DM_pgtz_Susceptibility}(a) and \ref{DM_pgtz_Susceptibility}(b).
For $D_p\neq 0$ and $D_z>0$, the increase of $D_z$ monotonically reduces 
$\chi_a$ from its value at $D_z=0$.
On the contrary, if $D_p\neq 0$ and $D_z<0$, a small increase of $|D_z|$
enhances $\chi_a$ at lower temperatures, producing a better agreement with 
the experiments than $\chi_a$ for $D_p\neq 0,\ D_z=0$. Ultimately, when 
$|D_z|\sim D_p$, the powder average of the susceptibility is again suppressed
with respect to its value for $D_z=0$. 

Based on the results presented in Fig.\ \ref{DM_pgtz_Susceptibility}, 
we conclude that the sharp upturn seen experimentally for the powder 
susceptibilities of the material \zncu\ can be understood to be a 
consequence of DM interactions for $D_p>|D_z|$ and $D_z<0$. We
predict that single crystal measurements should see an upturn in the 
anisotropy when the powder susceptibility departs from the KLHM result. 
Such a behavior has also been seen for spin-$\frac{5}{2}$ kagome system 
KFe$_3$(OH)$_6$(SO$_4$)$_2$.\cite{grohol05}

We should stress that in our theoretical calculations for the powder 
susceptibilities, we have assumed the $g$ factor to be isotropic. As 
depicted in Fig.\ \ref{DM_pgtz_Susceptibility}(d) for $D_z<0$, the 
$z$ susceptibility rises very rapidly, producing a large anisotropy 
$\chi_z/\chi_p$ [shown in the inset in 
Fig.\ \ref{DM_pgtz_Susceptibility}(d)]. Therefore, an expected 
anisotropic $g$ factor enhanced along $z$ will 
cause an even more rapid rise of $\chi_a$ than the one presented in 
the inset in Fig.\ \ref{DM_pgtz_Susceptibility}(b), and will lead to 
agreement with experiments with a smaller DM anisotropy. For example, 
already for $D_p=0.2J$ and $|D_z|<D_p$, the $z$ susceptibility is 
very similar to the experimental result for the powder averages.

Another factor that could reduce the DM anisotropy required to describe 
the experimental results is the presence of a small concentration 
of impurity spins.\cite{ran06,misguich07} However, as this may vary
from sample to sample, we will not consider it further in this work.
As mentioned in the Introduction, even in the absence of
single crystals, recent NMR work of Imai {\it et al.}\cite{imai07} 
may support the relevance of DM interaction to \zncu. The behavior 
they observe for the main peak and edges of the NMR spectra is 
similar to the powder [inset in Fig.\ \ref{DM_pgtz_Susceptibility}(c)] 
or $\chi_z$ [Fig.\ \ref{DM_pgtz_Susceptibility}(d)] and in-plane 
[Fig.\ \ref{DM_pgtz_Susceptibility}(c)] susceptibilities, 
respectively. 

\onecolumngrid

%%%%%%%%%%%%%%  FIGURE  %%%%%%%%%%%%%%%%%%%%%%%%%%%%%%%%%%%%%%%%%%%%%%
%%%%%%%%%%%%%%%%%%%%%%%%%%%%%%%%%%%%%%%%%%%%%%%%%%%%%%%%%%%%%%%%%%%%%%%%

\twocolumngrid

\subsection{Exchange anisotropies}

We discuss in what follows the role that exchange anisotropies
play in the behavior of the magnetic susceptibilities. 
Easy-plane ($\Delta>0$) and easy-axis ($\Delta<0$) exchange anisotropies 
can be introduced perturbing the Heisenberg Hamiltonian with a term 
\begin{equation}
{\cal H}_{EA}=\Delta \sum_{\langle i,j\rangle} 
\left( S_i^x S_j^x + S_i^y S_j^y \right).
\label{exchange}
\end{equation}

Equation (\ref{exchange}) breaks the SU(2) symmetry of the KLHM, however;
it does not couple different $S^z$ sectors. Hence, in the absence of 
$x$-$y$ magnetic fields, we can perform calculations for out-of-plane 
susceptibilities, entropy, and specific heat using the triangle-based NLC 
expansion summing contributions of up to eight triangles. Calculating the 
in-plane susceptibilities [Eq.\ (\ref{suscept})] requires introducing 
$x$-$y$ magnetic fields, which couples different $S^z$ sectors. Consequently, 
NLC calculations for $\chi_p$ can be done only up to six triangles. 
As we will show later, in the presence of exchange anisotropies, 
NLC results for $\chi_p$ are very similar to the ones obtained 
with ED (15 site cluster) down to $T\sim 0.35J$. For all other 
quantities, we only present NLC results, which are more accurate.

\subsubsection{Easy-plane exchange anisotropy}

\begin{figure}[!t]
\begin{center}
  \includegraphics[scale=.6,angle=0]{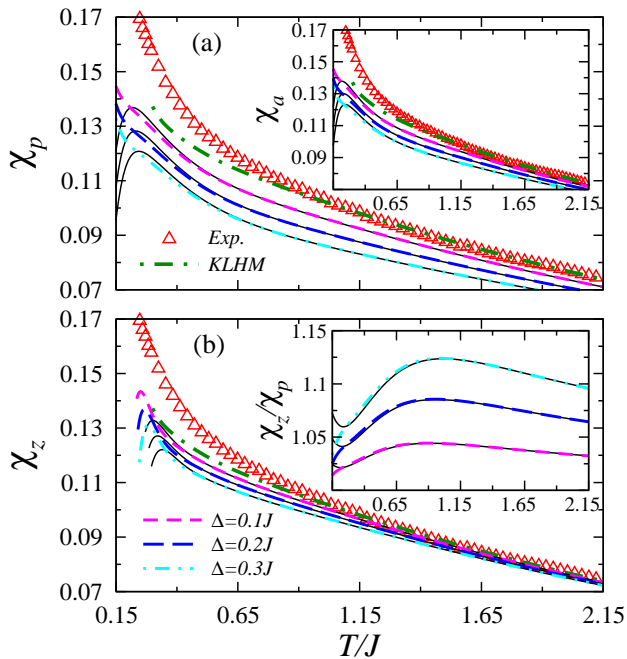}
\end{center}
\vspace{-0.5cm}
\caption{\label{XY_Susceptibility}
(Color online) Temperature dependence of the (a) in-plane and (b) out-of-plane 
susceptibilities in the presence of an easy-plane exchange anisotropy. Results 
are compared with $\chi$ for the pure KLHM (Ref.\ \onlinecite{nlc}) and with 
the experimental results (Refs.\ \onlinecite{helton07} and \onlinecite{ofer07}). 
In the inset in (a), we show the powder susceptibility, and in the inset in (b) 
the anisotropy. In (a) and the insets of (a) and (b), the susceptibilities 
were computed using ED. Thick (thin) lines show the ED results of the 15 (12) 
site cluster. In (b), the $z$ susceptibilities were obtained using NLC. 
Thick (thin) lines are the results of the NLC expansion with up to 
eight (seven) triangles.}
\end{figure}

Figure \ref{XY_Susceptibility} shows that an easy-plane exchange anisotropy 
decreases both the in-plane and out-of-plane susceptibilities with respect
to the pure Heisenberg model, and $\chi_z$ becomes larger than $\chi_p$, 
similar to the effect of a pure $D_z$ term shown in 
Fig.\ \ref{DM_z_Susceptibility}. There are, however, clear differences 
between $\Delta>0$ and $D_z\neq 0$. The easy-plane exchange anisotropy produces 
a large reduction of $\chi_p$ [Fig.\ \ref{XY_Susceptibility}(a)] and $\chi_a$ 
[inset in Fig.\ \ref{XY_Susceptibility}(a)] at high temperatures, and this 
reduction does not depend strongly on temperature down to $T\sim 0.3J$. 

%%%%%%%%%%%%%%  FIGURE  %%%%%%%%%%%%%%%%%%%%%%%%%%%%%%%%%%%%%%%%%%%%%%
%%%%%%%%%%%%%%%%%%%%%%%%%%%%%%%%%%%%%%%%%%%%%%%%%%%%%%%%%%%%%%%%%%%%%%%%

The above property of the exchange anisotropy highlights how remarkable 
DM interactions are. Their effect on the susceptibilities is negligible 
at high temperatures (even for large values of $D_z$) and only onsets 
as the temperature is lowered. Another important difference between an 
easy-plane exchange anisotropy and the pure out-of-plane DM term 
is that, as shown in the inset in Fig.\ \ref{XY_Susceptibility}(b), 
the behavior of the anisotropy $\chi_z/\chi_p$ produced by the former 
perturbation is nonmonotonic with temperature. Hence, susceptibility 
experiments with single crystals should be able to easily distinguish 
between these two types of anisotropies. 

\subsubsection{Easy-axis exchange anisotropy}

\begin{figure}[!t]
\begin{center}
  \includegraphics[scale=.6,angle=0]{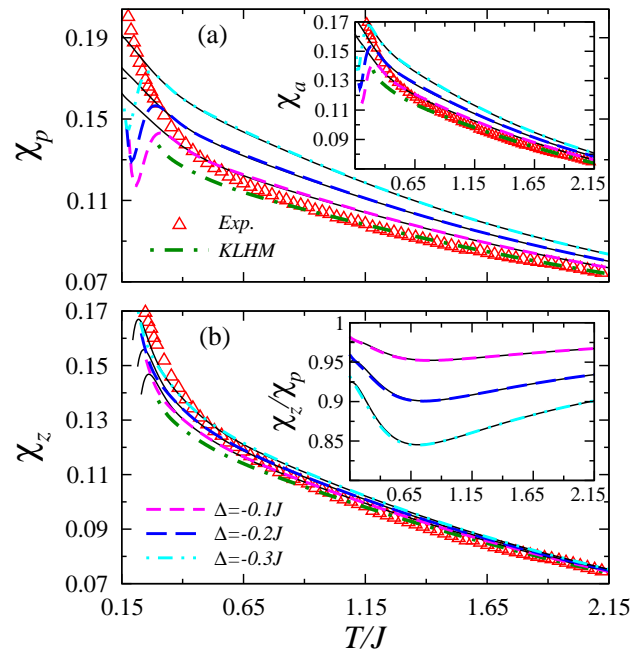}
\end{center}
\vspace{-0.5cm}
\caption{\label{IS_Susceptibility}
(Color online) Temperature dependence of the (a) in-plane and (b) out-of-plane 
susceptibilities in the presence of an easy-axis exchange anisotropy. Results 
are compared with $\chi$ for the pure KLHM (Ref.\ \onlinecite{nlc}) and with 
the experimental results (Refs.\ \onlinecite{helton07} and \onlinecite{ofer07}). 
In the inset in (a), we show the powder susceptibility, and in the inset in (b) 
the anisotropy. In (a) and its inset, we compare results of the NLC expansion 
with up to six triangles (thick lines) with ED of a 15 site cluster (thin lines). 
In (b), the susceptibilities were obtained using the NLC expansion with up 
to eight triangles (thick lines) and seven triangles (thin lines). 
The anisotropies [inset in (b)] were obtained using ED of 15 site 
(thick lines) and 12 site (thin lines) clusters.}
\end{figure}

The effect of an easy-axis exchange anisotropy on the susceptibility is 
depicted in Fig.\ \ref{IS_Susceptibility}. Like $D_p$, $\Delta<0$ enhances 
both $\chi_p$ and $\chi_z$ with respect to $\chi$ in the pure KLHM case.
However, $\Delta<0$ has a large effect on $\chi_p$ 
[Fig.\ \ref{IS_Susceptibility}(a)] and $\chi_a$ 
[inset in Fig.\ \ref{IS_Susceptibility}(a)] at high temperatures. 
Basically, $\Delta<0$ produces a large enhancement of the susceptibility 
that is not strongly dependent on the temperature down to 
$T\sim 0.3J$.\cite{nlcvsed} Hence, $\Delta<0$ 
cannot provide an explanation for the sharp increase seen experimentally 
in the powder susceptibility. Notice also that in presence 
of an easy-axis exchange anisotropy, the in-plane susceptibilities are 
larger than the out-of-plane ones, and their anisotropy [inset in 
Fig.\ \ref{IS_Susceptibility}(b)] is nonmonotonic in temperature.

%%%%%%%%%%%%%%  FIGURE  %%%%%%%%%%%%%%%%%%%%%%%%%%%%%%%%%%%%%%%%%%%%%%
%%%%%%%%%%%%%%%%%%%%%%%%%%%%%%%%%%%%%%%%%%%%%%%%%%%%%%%%%%%%%%%%%%%%%%%%

\section{Entropy and Specific heat \label{entropycv}}

In this section, we to study the entropy ($S$),
\begin{equation}
S=\frac{1}{N} (\ln\,Z +\langle{\cal H}\rangle/T),
\label{entropy}
\end{equation}
and specific heat ($C_v$),
\begin{eqnarray}
C_v= \frac{1}{N T^2} \left(\langle{\cal H}^2\rangle-
                           \langle{\cal H}  \rangle^2\right) ,
\label{cv}
\end{eqnarray}
of the KLHM in the presence of various perturbations.
In Eqs.\ (\ref{entropy}) and (\ref{cv}), $N$ 
stands for the number of lattice sites and $Z$ for the partition 
function.

\subsection{$D_z\neq0,\ D_p=0$, and $\Delta>0$}

\begin{figure}[!h]
\begin{center}
  \includegraphics[scale=.6,angle=0]{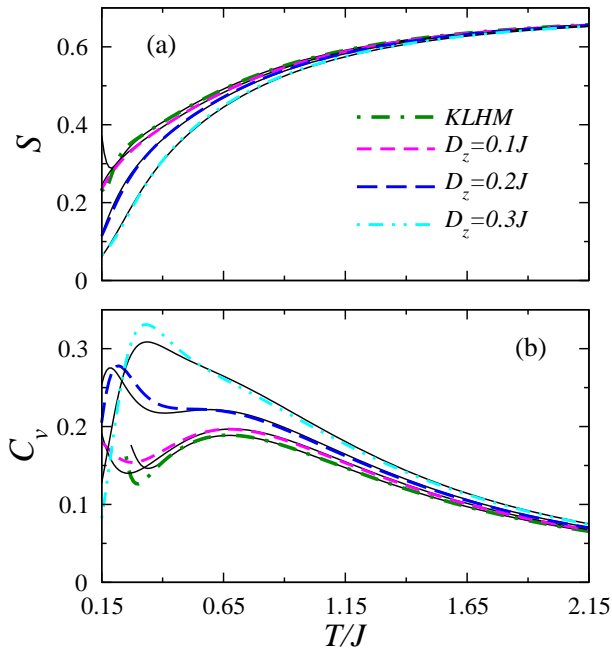}
\end{center}
\vspace{-0.5cm}
\caption{\label{DM_z_EntropyCv}
(Color online) (a) Entropy and (b) specific heat as a function of the 
temperature in the presence of an out-of-plane DM interaction. 
For the KLHM, we used the NLC triangle-based expansion (Ref.\ \onlinecite{nlc}) 
considering up to eight triangles (thick line) and seven triangles (thin line). 
For all plots with $D_z\neq 0$, thick (thin) lines show the ED results 
of the 15 (12) site cluster calculation.
}
\end{figure}

We start by considering the effect of an out-of-plane DM anisotropy.
Results for $S$ and $C_v$ and different values of $D_z$ are shown in 
Fig.\ \ref{DM_z_EntropyCv}. As expected from the qualitative analysis
in the previous section, where we argued that $D_z$ breaks the 
degeneracy among different possible N\'eel states for KLHM, 
Fig.\ \ref{DM_z_EntropyCv}(a) shows that as the temperature is 
reduced, $D_z$ reduces the entropy with respect to its value for the KLHM.
This suppression of the entropy is accompanied by a large increase in
the specific heat with respect to the KLHM. While for very small 
values of $D_z$ a high-temperature peak can still be seen in $C_v$ 
[Fig.\ \ref{DM_z_EntropyCv}(b)], by the time $D_z=0.3J$, any evidence 
of such a peak has disappeared. 

%%%%%%%%%%%%%%  FIGURE  %%%%%%%%%%%%%%%%%%%%%%%%%%%%%%%%%%%%%%%%%%%%%%
%%%%%%%%%%%%%%%%%%%%%%%%%%%%%%%%%%%%%%%%%%%%%%%%%%%%%%%%%%%%%%%%%%%%%%%

\begin{figure}[!h]
\begin{center}
  \includegraphics[scale=.6,angle=0]{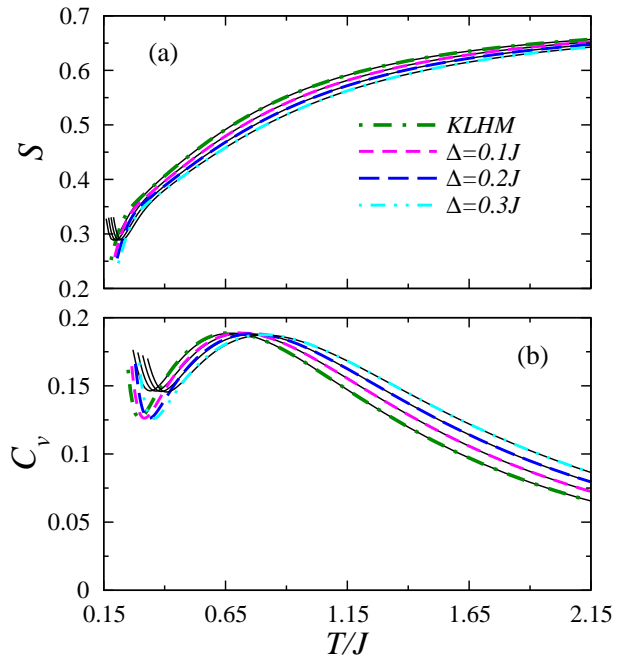}
\end{center}
\vspace{-0.5cm}
\caption{\label{XY_EntropyCv}
(Color online) (a) Entropy and (b) specific heat as a function of the 
temperature in the presence of an easy-plane exchange anisotropy. 
All results were obtained using the NLC triangle-based expansion 
(Ref.\ \onlinecite{nlc}) considering up to eight triangles (thick line) 
and seven triangles (thin line).}
\end{figure}

As mentioned in Sec.\ \ref{susceptibility}, the easy-plane exchange 
anisotropy [Eq.\ (\ref{exchange})] produces some features that are 
similar to a pure $D_z$ term. In Fig.\ \ref{XY_EntropyCv}, we show 
the effects of $\Delta>0$ on $S$ and $C_v$. Indeed, like $D_z$, 
an easy-plane anisotropy suppresses the entropy with respect to the 
KLHM. This reduction, however, is apparent at high temperatures 
even if $\Delta$ is small, while in the presence of $D_z$ 
[Fig.\ \ref{DM_z_EntropyCv}(a)], it is only noticeable as the 
temperature is lowered. 

The effect of $\Delta>0$ on the specific heat is very 
different from that of $D_z$, as seen in Fig.\ \ref{XY_EntropyCv}(b). 
Increasing $\Delta$ (up to $\Delta=0.3J$) only displaces the high-temperature 
peak to higher temperatures almost without changing its height. (Eventually, 
as the exchange anisotropy is further increased toward the $XY$ limit, the 
height of the $C_v$ peak would decrease.) This means that if one could
extract the behavior of $S$ and $C_v$ from experiments, one could further
distinguish between DM and exchange anisotropies.

%%%%%%%%%%%%%%  FIGURE  %%%%%%%%%%%%%%%%%%%%%%%%%%%%%%%%%%%%%%%%%%%%%%
%%%%%%%%%%%%%%%%%%%%%%%%%%%%%%%%%%%%%%%%%%%%%%%%%%%%%%%%%%%%%%%%%%%%%%%

\subsection{$D_p\neq0,\ D_z=0$, and $\Delta<0$}

\begin{figure}[!h]
\begin{center}
  \includegraphics[scale=.6,angle=0]{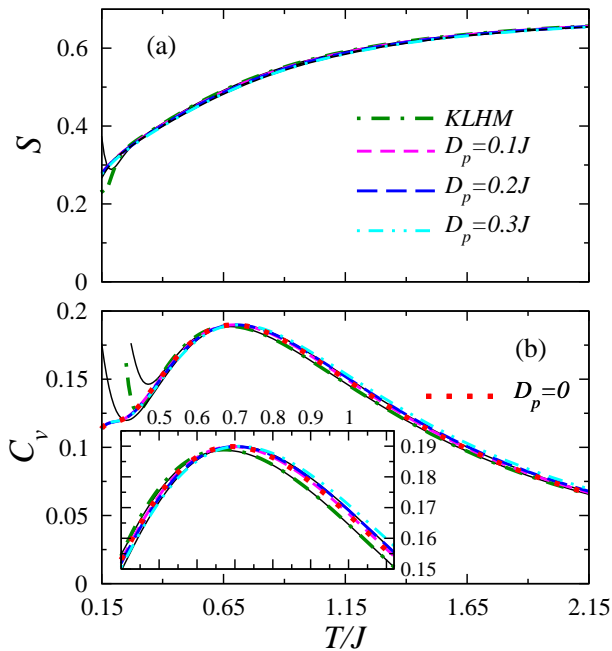}
\end{center}
\vspace{-0.5cm}
\caption{\label{DM_p_EntropyCv}
(Color online) (a) Entropy and (b) specific heat as a function of the 
temperature in the presence of an in-plane DM interaction. 
For the KLHM, we used the NLC triangle-based expansion (Ref.\ \onlinecite{nlc}) 
considering up to eight triangles (thick line) and seven triangles (thin line). 
For all plots with $D_p\neq 0$, thick (thin) lines show the ED results of 
the 15 (12) site cluster calculation. In (b), the extra plot denoted 
by $D_p=0$ is the KLHM ED result obtained with a 15 site cluster. The
inset in (b) magnifies the high-temperature peak of $C_v$ so that 
ED finite-size effects become discernible.
}
\end{figure}

We now turn to the effects of a pure in-plane DM anisotropy on the 
entropy and specific heat. This is depicted in Fig.\ \ref{DM_p_EntropyCv}. 
Figures \ref{DM_p_EntropyCv}(a) and \ref{DM_p_EntropyCv}(b) show that, at 
least down to temperatures $\sim$$0.15J$, $D_p$ has a negligible effect on 
the entropy and specific heat, respectively. Considering that the largest 
$D_p$ in Fig.\ \ref{DM_p_EntropyCv} is 30\% of $J$ and that such 
anisotropy produces large changes in the uniform susceptibilities 
(Fig.\ \ref{DM_p_Susceptibility}), we find this feature remarkable.

%%%%%%%%%%%%%%  FIGURE  %%%%%%%%%%%%%%%%%%%%%%%%%%%%%%%%%%%%%%%%%%%%%%
%%%%%%%%%%%%%%%%%%%%%%%%%%%%%%%%%%%%%%%%%%%%%%%%%%%%%%%%%%%%%%%%%%%%%%

We should add that as $D_p$ is increased, the small displacement seen 
in the high temperature peak [Fig.\ \ref{DM_p_EntropyCv}(b)] toward higher 
temperatures is of the same order as the ED finite-size effects. To make 
that clear, we have also plotted $D_p=0$ results obtained from the exact 
diagonalization of a finite cluster with 15 sites. As better seen in the 
inset, the ED peak for $D_p=0$ is slightly displaced toward higher 
temperatures than the thermodynamic limit result provided by NLC, and that 
displacement is of the same order as the one seen for $D_p\neq0$. 
($C_v$ is, in general, very sensitive to finite-size effects.\cite{nlc})

\begin{figure}[!h]
\begin{center}
  \includegraphics[scale=.6,angle=0]{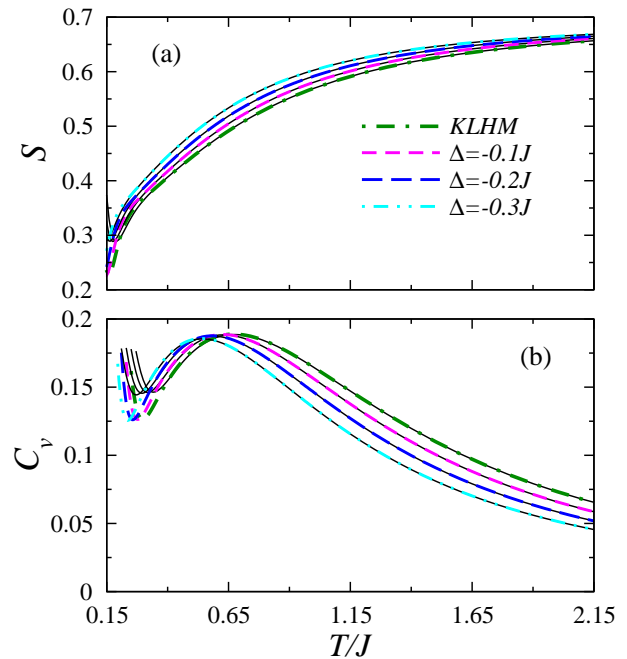}
\end{center}
\vspace{-0.5cm}
\caption{\label{IS_EntropyCv}
(Color online) (a) Entropy and (b) specific heat as a function of the 
temperature in the presence of an easy-axis exchange anisotropy. 
All results were obtained using the NLC triangle-based expansion
(Ref.\ \onlinecite{nlc}) considering up to eight triangles (thick line) 
and seven triangles (thin line).}
\end{figure}

The presence of an easy-axis exchange anisotropy has a very different 
effect on $S$ and $C_v$ of the KLHM. This can be seen by comparing 
Fig.\ \ref{IS_EntropyCv} with Fig.\ \ref{DM_p_EntropyCv}. The increase 
of $|\Delta|$ increases the entropy, at all temperatures, with respect 
to the KLHM. This is expected since in the Ising limit, the system has 
a finite entropy at zero temperature. For $C_v$, what happens is that 
the high-temperature peak moves to lower temperatures. The height of 
the peak almost does not change up to $\Delta=-0.3J$, although it 
eventually decreases as $\Delta$ approaches -1.\cite{nlc}

%%%%%%%%%%%%%%  FIGURE  %%%%%%%%%%%%%%%%%%%%%%%%%%%%%%%%%%%%%%%%%%%%%%
%%%%%%%%%%%%%%%%%%%%%%%%%%%%%%%%%%%%%%%%%%%%%%%%%%%%%%%%%%%%%%%%%%%%%%%

\subsection{$D_p\neq0,\ D_z\neq0$}

If both DM terms are present in the system, we find that the 
deviations of $S$ and $C_v$ from the KLHM result (in the range of 
temperatures discussed in this work) are mainly determined by the
value of $D_z$, almost independent of the value (and sign)
of $D_p$ (up to $D_p\sim 0.3J$). Hence, at least at intermediate 
and high temperatures, $S$ and $C_v$ are quite insensitive to the 
existence of an in-plane DM interaction. This is further discussed
in a later section in comparison to experiments on \zncu.

\subsection{Quenched dilution}

As discussed in Ref.\ \onlinecite{rigol07}, another important perturbation 
of the KLHM is the presence of quenched dilution. Quenched dilution 
could be generated in the materials \zncu\ due to the substitution of 
Cu sites in the kagome planes by Zn. The missing spins on the lattice 
could create local moments in the singlet background and cause a Curie-like 
susceptibility to arise as the temperature is lowered. However, we have 
shown that at least down to temperatures $T\sim 0.3J$, the only effect 
that such a dilution has on the susceptibility is to suppress it with 
respect to the KLHM result.\cite{rigol07} In this sense, quenched 
dilution has the opposite effect of impurity spins as the latter enhance 
the susceptibility.

Here, we present studies of the effects of quenched dilution 
on entropy and specific heat of KLHM. Our calculations are
performed using the triangle-based NLC expansion considering up to 
eight triangles. If $c$ is the dilution, we assume that at each site we 
have a hole with probability $c$ and a spin with probability $1-c$. 
The holes are fixed in their position and extensive quantities are 
averaged over all possible configurations $C$ using the relation
\begin{equation}
\langle O\rangle=\sum_{C} P(C) O(C),
\end{equation}
where 
\begin{equation}
 P(C)= c^{N_h} (1-c)^{N_s}
\end{equation}
is the probability of the configuration $C$ with $N_h$ holes and
$N_s$ spins. 

%%%%%%%%%%%%%%  FIGURE  %%%%%%%%%%%%%%%%%%%%%%%%%%%%%%%%%%%%%%%%%%%%%%
\begin{figure}[!h]
\begin{center}
  \includegraphics[scale=.58,angle=0]{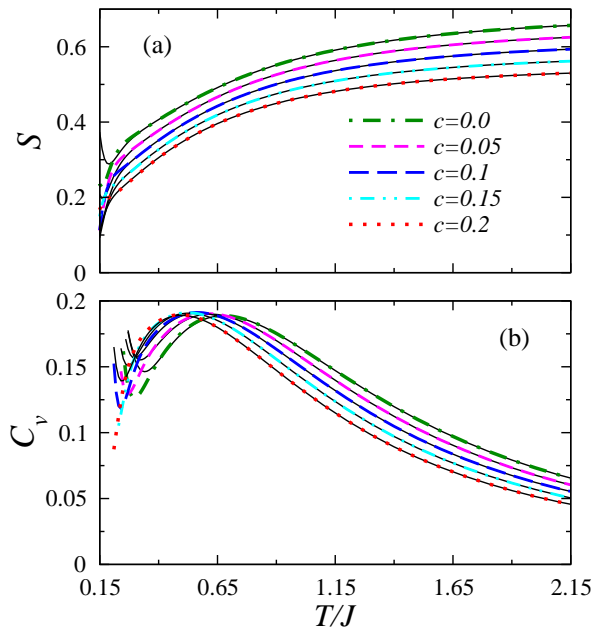}
\end{center}
\vspace{-0.5cm}
\caption{\label{QH_EntropyCv}
(Color online) (a) Entropy and (b) specific heat as a function of the 
temperature in the presence of quenched dilution. 
All results were obtained using the NLC triangle-based expansion
(Ref.\ \onlinecite{nlc}) considering up to eight triangles (thick line) 
and seven triangles (thin line).}
\end{figure}
%%%%%%%%%%%%%%%%%%%%%%%%%%%%%%%%%%%%%%%%%%%%%%%%%%%%%%%%%%%%%%%%%%%%%%%

The entropy and specific heat for several values of hole concentration 
are shown in Fig.\ \ref{QH_EntropyCv}. We note that at these 
intermediate temperatures [Fig.\ \ref{QH_EntropyCv}(a)], holes simply 
lower the entropy at all temperatures. In the case of the specific heat,
they just displace the high temperature peak to lower temperatures, 
almost without changing its height.

\subsection{Entropy difference from kagome lattice Heisenberg model 
and implication for \zncu}

\begin{figure}[!h]
\begin{center}
  \includegraphics[scale=.63,angle=0]{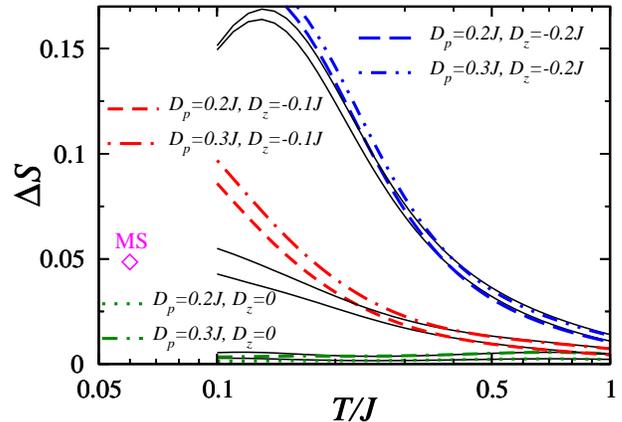}
\end{center}
\vspace{-0.5cm}
\caption{\label{Exp_Entropy}
(Color online) Difference between the entropy of the pure KLHM and 
the entropy of the KLHM in the presence of DM anisotropies. 
Thick dashed lines show the ED results of the 15 site cluster and
the thin black lines result of 12 site cluster.
The diamond at $T=0.06J$ depicts the minimum difference between 
the pure KLHM and the experimental result as obtained by MS in 
Ref.\ \onlinecite{misguich07}.}
\end{figure}

To conclude the section on specific heat and entropy, we study the
entropy difference between the KLHM and the model with different 
DM parameters. Shown in Fig.\ \ref{Exp_Entropy} is $\Delta S$, 
given by
\[
\Delta S=S(D_p=0,D_z=0)-S(D_p,D_z).
\]
It is the reduction in entropy due to DM interactions. In 
Fig.\ \ref{Exp_Entropy}, one notices that the entropy reduction 
is determined primarily by $D_z$ and $D_p$ plays a small role. 

The diamond in Fig.\ \ref{Exp_Entropy} represents the minimum 
discrepancy between the pure KLHM and \zncu\ at $T/J=0.06$ as 
determined by Misguich and Sindzingre (MS) in Ref.\ \onlinecite{misguich07}. 
The experimental data at higher temperatures are likely dominated by phonons. 
From Fig.\ \ref{Exp_Entropy}, we conclude that $|D_z|/J$ is likely to be 
about $0.1$ in \zncu. Hence, while the sharp increase in 
the susceptibility discussed in Sec.\ \ref{susceptibility} allows 
us to make an estimate of the possible values of $D_p$ 
($D_p/J\approx 0.2-0.3$), the entropy reduction allows us to get 
an estimate of $|D_z|$.

%%%%%%%%%%%%%%  FIGURE  %%%%%%%%%%%%%%%%%%%%%%%%%%%%%%%%%%%%%%%%%%%%%%
%%%%%%%%%%%%%%%%%%%%%%%%%%%%%%%%%%%%%%%%%%%%%%%%%%%%%%%%%%%%%%%%%%%%%%%

\section{Conclusions \label{conclusions}}

In this work, we have studied in detail the effect of perturbations
such as in-plane and out-of-plane DM interactions, exchange 
anisotropies, and quenched dilution on the KLHM. We have focused
here on the effect such perturbations have on
magnetic susceptibilities, entropy, 
and specific heat. We first summarize our theoretical findings:

(i) In the presence of a pure out-of-plane DM term ($D_z\neq0,\ D_p=0$)
both in-plane ($\chi_p$) and out-of-plane ($\chi_z$) susceptibilities
are suppressed with respect to the KLHM result, and $\chi_z$ becomes
larger than $\chi_p$. However, the susceptibility anisotropy 
$\chi_z/\chi_p$ is not large ($<1.1$) down to $T\sim 0.5J$
for $D_z\lesssim 0.3J$. On the other hand,
a $D_z$ term suppresses the entropy as the 
temperature is lowered and produces an increase of the specific 
heat at intermediate temperatures.

(ii) In the presence of a pure in-plane DM term ($D_z=0,\ D_p\neq0$),
both in-plane ($\chi_p$) and out-of-plane ($\chi_z$) susceptibilities
are enhanced with respect to the KLHM result and, as for pure $D_z$, 
$\chi_z$ becomes larger than $\chi_p$. The susceptibility anisotropy 
is not large ($<1.1$) down to $T\sim 0.25J$ for $D_p\lesssim 0.3J$. 
A $D_p$ term ($\lesssim 0.3J$) has a negligible influence on 
entropy and specific heat down to $T\sim 0.15J$.

(iii) When both $D_p$ and $D_z$ are present, the susceptibility becomes
sensitive to the sign of $D_z$. (A) For a constant value of $D_p$, 
the increase of $D_z$ for $D_z>0$ suppresses both the in-plane and 
out-of-plane susceptibilities with respect to the $D_z=0$ value. 
The susceptibility anisotropy is of the same order as when only
$D_p$ or $D_z$ is present. (B) Also keeping $D_p$ constant, the 
increase of $|D_z|$ for $D_z<0$ suppresses the in-plane susceptibility 
but enhances the out-of-plane susceptibility. 

(iv) In the presence of both $D_p$ and $D_z$, the entropy and 
specific heat are mainly determined by the modulus of $D_z$. $D_p$ and 
the sign of $D_z$ play a relatively small role.

(v) The presence of an easy-plane exchange anisotropy
suppresses both $\chi_p$ and $\chi_z$
from their KLHM values, and $\chi_z$ becomes larger than $\chi_p$. 
Such anisotropy has a large effect even at high temperatures 
and does not produce deviations from the KLHM that are 
strongly temperature dependent. In addition, 
it generates susceptibility anisotropies 
that are nonmonotonic and weakly dependent on temperature 
down to $T\sim 0.25J$. $\Delta>0$ also suppresses the entropy
with respect to the KLHM. The high-temperature peak of the 
specific heat is slightly displaced toward higher temperatures almost 
without modifying its height (for $\Delta\lesssim0.3J$).

(vi) The presence of an easy-axis exchange anisotropy 
enhances both $\chi_p$ and $\chi_z$
from their KLHM values, and $\chi_p$ becomes larger than $\chi_z$. 
$\Delta<0$ has a large effect on the high-temperature values
of $\chi_p$ and also does not produce deviations from the KLHM that are 
strongly temperature dependent. It also leads to susceptibility 
anisotropies that are nonmonotonic and weakly dependent on the temperature 
down to $T\sim 0.25J$. $\Delta<0$ also enhances the entropy 
with respect to the KLHM result. The high-temperature peak of the specific 
heat is slightly displaced toward lower temperatures without much change
in height (for $|\Delta|\lesssim0.3J$).

(vii) At intermediate and high temperatures ($T\gtrsim 0.3J$), quenched 
dilution has been shown to suppress the uniform susceptibility with 
respect to the KLHM.\cite{rigol07} We have discussed here that it
also reduces the entropy for all temperatures $T\gtrsim 0.3J$. In the 
case of the specific heat, the effect of dilution is to displace the 
high temperature peak toward lower temperatures without affecting 
its height (at least when $c\lesssim 0.2$).

We now discuss our conclusions with regard to the observed properties
of the material \zncu: 

(i) The observed susceptibility shows large
enhancement with respect to the KLHM, which has a sudden onset
below $T=J/2$. This kind of behavior is only compatible, within
the models studied, with a large $D_p\approx 0.2-0.3$ and a $D_z<0$,
with $|D_z|<D_p$.

(ii) Misguich and Sindzingre\cite{misguich07} have previously 
concluded that the experiments on \zncu\  show that there is a 
large reduction in entropy with respect to KLHM at $T/J=0.06$. 
This reduction is at least $0.05$ and may be larger when impurities 
and phonons are taken into account. The comparison with DM anisotropy 
calculations shows that this implies $|D_z|/J\approx 0.1$.

Based on these results, our overall conclusion for the parameters of the
material \zncu\ is $D_p/J$ in the range $0.20-0.30$, $|D_z|/J\approx 0.1$,
and $J\approx 170K$. All these numbers could change if there are substantial 
impurity contributions present. Future experiments on the effects of 
anisotropy can resolve these issues. 

On the theoretical side, an important question that remains open is the nature 
of the low-temperature phase(s) of the KLHM in the presence of DM terms. 
In classical systems, it has been shown that ground state is 
ordered,\cite{elhajal02,harris06} and there is a finite 
temperature phase transition in which the critical temperature 
depends on the values of $D_z$ and $D_p$.\cite{elhajal02} However, one should 
keep in mind that for classical systems, even in the absence of DM anisotropy, 
the system orders as $T$ goes to zero.\cite{huse,reimers93} Quantum effects 
for the spin-$\frac{1}{2}$ case, and their relation to the experimental 
absence of any order down to 50 mK, still need to be elucidated.

\begin{acknowledgments}

This work was supported by the US National Science Foundation, Grants 
Nos.\ DMR-0240918, DMR-0312261, and PHY-0301052. We are grateful to Oren Ofer, 
Amit Keren, Joel Helton, and Young Lee for providing us with the experimental 
susceptibility data, and to Michael Hermele and Takashi Imai for valuable 
discussions. We thank the HPCC-USC center where all our computations
have been performed. \\

\end{acknowledgments}

{\it Note added on proof.} --It has been argued in recent experiments
\cite{vries07,lee07,bert07} that contrary to the original expectation 
a rather large concentration of intersite mixing (Cu/Zn) impurities 
($c_s\sim6-10$\%) may be present in the Herbertsmithite material. 
One natural question that has not been addressed in these works is 
whether such a large impurity concentration can be accommodated to 
reproduce the intermediate and high temperature magnetic susceptibility 
of the KLHM. In Ref.\ \onlinecite{rigol07} we have shown that subtracting 
the contribution of $c_s=4.5$\% free impurity spins ($J=200$ K) to the experimental 
results of Refs.\ \onlinecite{helton07} and \onlinecite{ofer07} one can reproduce 
the magnetic susceptibility of the KLHM down to $T=0.3J$. (The free 
impurity spin concentration reported in Ref.\ \onlinecite{rigol07} 
was incorrect by a factor 3/2, which means that in Fig.\ 2 of that reference 
one should read $c=0.045$ and $c=0.09$ instead of $c=0.03$ and $c=0.06$, 
respectively.) However, a sharp rise in the 
susceptibility remained below $T=0.3J$ that was not expected in the KLHM. 
It was shown later by Misguich and Sindzingre\cite{misguich07} that adding 
a small ferromagnetic coupling between impurities ($c_s=3.7$\% and $J=190$ K) 
allows one to reproduce the magnetic susceptibility of the KLHM down to 
$T=0.2J$, with $\chi$ starting to decrease when $T\sim 0.1J$.

\begin{figure}[!h]
\begin{center}
  \includegraphics[scale=.63,angle=0]{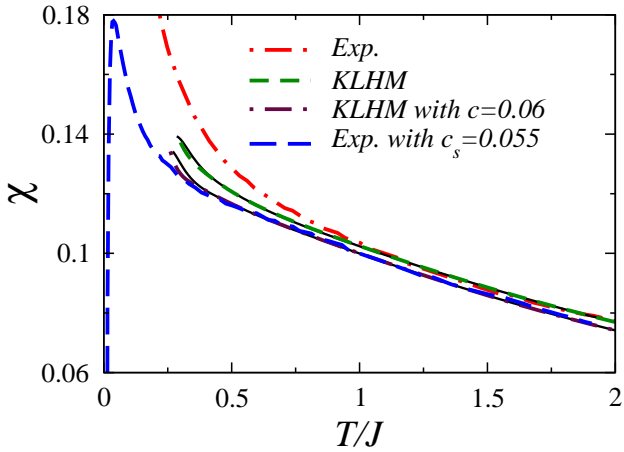}
\end{center}
\vspace{-0.5cm}
\caption{\label{SpinImpurities}
(Color online) NLC results for the KLHM with ($c=6$\%) and 
without ($c=0$) nonmagnetic impurities are compared with the 
experimental results after a $c_s=5.5$\% and $c_s=0$ contribution 
of free impurity spins is subtracted, i.e., in the latter case 
we have plotted $\chi_{molar}/C-c_s/(4T)$ (where $C$ was defined in 
Sec.\ \ref{susceptibility}). In the presence of impurities we
obtain $J=210$ K, $g=2.37$ as opposed to $J=170$ K, $g=2.33$
in their absence. NLC results for the eight(seven) triangle
based expansion are plotted as thick(thin) lines.}
\end{figure}

From the theoretical results in Refs.\ \onlinecite{rigol07} and \onlinecite{misguich07} 
one could conclude that after subtracting the large contribution of $c_s\sim6-10$\% 
impurity spins (Cu) to the experimental results one would obtain a magnetic susceptibility 
that is incompatible with the KLHM at intermediate and high 
temperatures. However, such a large concentration of impurity spins can be 
accommodated if one considers the effect of the non-magnetic impurity 
counterpart (Zn) that is present in the kagome planes. As shown in 
Ref.\ \onlinecite{rigol07} such impurities reduce the magnetic 
susceptibility with respect to the KLHM. Figure \ref{SpinImpurities}
shows that after subtracting a $c_s\sim 6$\% free impurity spin contribution 
from the experimental results one can reproduce the susceptibility of the 
KLHM with a $c=6$\% of quenched nonmagnetic impurities.

%%%%%%%%%%%%%%  FIGURE  %%%%%%%%%%%%%%%%%%%%%%%%%%%%%%%%%%%%%%%%%%%%%%
%%%%%%%%%%%%%%%%%%%%%%%%%%%%%%%%%%%%%%%%%%%%%%%%%%%%%%%%%%%%%%%%%%%%%%%

The results presented in Fig.\ \ref{SpinImpurities} show that a large 
concentration of intersite mixing (Cu/Zn) impurities is compatible 
with the KLHM at intermediate and high temperatures. However, whether 
impurities are the main contribution to the magnetic susceptibility 
at intermediate and low temperatures, as opposed to DM interactions, 
still needs to be clarified measuring the anisotropy in the in-plane 
and out-of-plane susceptibilities for temperatures where $\chi$ clearly 
departs from the pure KLHM result.

\appendix

\section{Dzyaloshinsky-Moriya interactions under a different lattice 
symmetry}

%%%%%%%%%%%%%%  FIGURE  %%%%%%%%%%%%%%%%%%%%%%%%%%%%%%%%%%%%%%%%%%%%%%
\begin{figure}[t]
\begin{center}
  \includegraphics[scale=.6,angle=0]{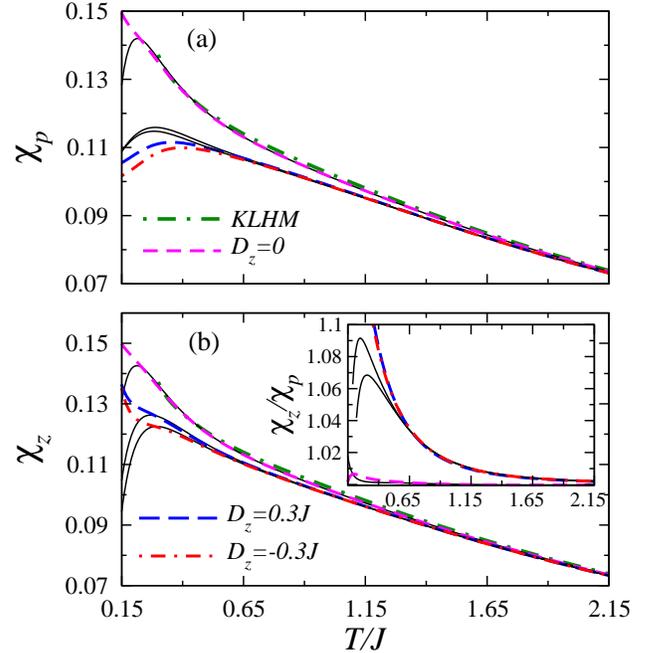}
\end{center}
\vspace{-0.5cm}
\caption{\label{DM_NotExp_p_Susceptibility}
(Color online) Temperature dependence of the (a) in-plane and (b) out-of-plane 
susceptibilities in the presence of a $D_p=0.3J$ anisotropy (with ${\bf D}_p$ 
alternating from triangle to triangle as explained in the text) and different 
values of $D_z$. The results are compared with $\chi$ for the pure KLHM 
(Ref.\ \onlinecite{nlc}) In the inset in (b), we show the anisotropy produced 
by these DM terms. In all plots of DM calculations, thick (thin) lines show 
the ED results of the 15 (12) site cluster.}
\end{figure}
%%%%%%%%%%%%%%%%%%%%%%%%%%%%%%%%%%%%%%%%%%%%%%%%%%%%%%%%%%%%%%%%%%%%%%%%

In Sec.\ \ref{susceptibility}, we discuss the effects of the DM anisotropy allowed
by the symmetry of \zncu\ on the susceptibility of the KLHM. In \zncu, the bonds 
between Cu$^{2+}$ ions 
(which have an oxygen atom in the middle) are distorted away from the kagome
planes, and the direction of this distortion alternates from triangle to triangle.
This symmetry of \zncu\ sets the direction of ${\bf D}_p$, which within our 
notation is to be always pointing toward the center of the 
triangles.\cite{elhajal02,harris06} [In Fig.\ \ref{DM_z_Susceptibility}, 
the ${\bf D}_p$ that multiplies $({\bf S_1}\times{\bf S_2})_y$ is (0,$D_p$,0), 
and the one that multiplies $({\bf S_4}\times{\bf S_5})_y$ is (0,$-D_p$,0).] 
If the bonds between the Cu sites would not be distorted at all, a perfect 
kagome lattice would be embedded in three dimensions, because of the symmetry of 
the lattice ${\bf D}_p=0$, and only $D_z$ could be different from 
zero.\cite{dzyaloshinsky58,moriya60}

For completeness, we briefly discuss here how the susceptibility of the 
KLHM would behave in a material where the bonds between magnetic ions 
are all distorted in the same direction away from the kagome planes. In this 
case, the $D_z$ terms in Eq.\ (\ref{DMH}) are identical to the ones we have 
considered for \zncu, but the ${\bf D}_p$ vectors will, within our notation, 
alternate pointing inward and outward of the up-pointing and down-pointing 
triangles, respectively. That scenario seems to be very interesting within the 
discussion in Ref.\ \onlinecite{ran06}.\cite{michael}

In Fig.\ \ref{DM_NotExp_p_Susceptibility}, we depict the $x$-$y$ and $z$ 
susceptibilities as a function of the temperature for $D_p=0.3J$ and 
three different values of $D_z$. One can see in 
Fig.\ \ref{DM_NotExp_p_Susceptibility} that, for $D_z=0$, the kind of 
$D_p$ anisotropy discussed in this appendix has almost no effect 
on the susceptibility of the KLHM, at least for the temperatures 
considered in this work. The inset in (b) shows that it also does not 
generate any asymmetry between $\chi_p$ and $\chi_z$. Introducing a 
finite $D_z$, either positive or negative, only suppresses (in a very 
similar way independent of the sign) both in-plane and out-of-plane 
susceptibilities with respect to the KLHM. (Similar to the discussion 
in Sec.\ \ref{entropycv}, the $D_p$ anisotropy discussed in this 
appendix is almost irrelevant to the entropy and specific heat.)

Hence, in kagome lattice materials with a crystal symmetry different 
from that of \zncu, $D_p$ anisotropies will produce a very different behavior 
of the susceptibility, which will not be enhanced with respect to the 
one of the KLHM.


\begin{thebibliography}{99}
\bibitem{experiments} 
For both experimental cases, we have taken $J=170$ K. The $g$ factors 
needed for the fit are 2.33 for the data from Ref.\ \onlinecite{helton07} 
and 2.19 for the data from Ref.\ \onlinecite{ofer07}.
\bibitem{sign} 
Thermodynamic quantities studied in this work are unaffected by the sign 
of $D_p$. We assume it to be positive without losing generality.
\bibitem{nlcvsed} 
Notice that the NLC results for $\chi_p$ and 
$\chi_a$ up to six triangles are almost indistinguishable from the ED 
result of a $15$ site cluster down to $T\sim 0.35J$.
\end{thebibliography}
\end{document}